# A comment on Eta Carinae's Homunculus Nebula imaging

Amelia Carolina Sparavigna
Department of Physics, Politecnico di Torino, Torino, Italy

**Abstract**
Homunculus Nebula is surrounding the star system Eta Carinae. The nebula is embedded within a much larger ionized hydrogen region, which is the Carina Nebula. Homunculus is believed to have been ejected in a huge outburst from Eta Carinae in 1841, so brightly to be visible from Earth. This massive explosion produced two polar lobes and an equatorial disc, moving outwards. Though Eta Carinae is quite away, approximately 7,500 light-years, it is possible to distinguish in the nebula, many structures with the size of about the diameter of our solar system. Knots, dust lanes and radial streaks appear quite clearly in many images. In this paper, we compare the imaging of Homunculus Nebula has obtained by HST and Gemini South Telescope research teams. We use some processing methods, to enhance some features of the structure, such as the color gradient, and knots and filaments in the central part of the nebula.

**Introduction**
Space telescopes and Earth-based telescopes with adaptive optics provide a huge amount of data, that after a subsequent image processing, are submitted for scientific analysis. Using a combination of many processing techniques, often including deconvolution methods, researchers are creating very high resolution images of extended objects such as nebulae. These images reveals so many details, that the researchers can try to model the history of the nebula expansion. Even faint structures such as the bow shocks created by stellar winds can appear in these images [1]. Let us remember that the images we can see in the scientific literature and published on the world wide web sites, are coming not only from an observation in the visible range of radiation, but are often generated with filters at several specific wavelengths. Resulting images are then composed with a superposition of signals ranging from the infrared to the ultraviolet radiation. This spanning of a large interval of wavelength has many advantages. The infrared astronomy for example, with data coming from devices equipped with infrared sensors, is able to penetrate the dusty regions of space, such as the molecular clouds in nebulae, and detect the planets revolting about our neighboring stars [2,3]. Moreover, the use of filters gives information on the temperature of the observed structures.
Here we discuss the benefit of a further image processing of astronomical images in enhancing specific details in the images. The processing will be applied on images of a specific object, the Homunculus Nebula of Eta Carinae. This nebula is embedded within a much larger ionized hydrogen region, which is the Carina Nebula. The Homunculus is believed to have been ejected in an outburst in 1841, so brightly to be visible from Earth. This massive explosion produced two polar lobes and an equatorial disk, which are moving outwards from the star [4,5]. We will show processed images, from the originals by the Hubble Space Telescope (HST) and the Gemini South Telescope in Chile, to enhance some features of the structure, such as the color gradient, and the knots and laces in the central part of the nebula. The processing allows an interesting comparison of structures in both images. It helps identifying some knots, the positions of which could be used as reference points for quantitative evaluations.

**Eta Carinae and her little man**
Eta Carinae's Homunculus (little man in Latin) is a bipolar nebula, where we see a pair of roughly spherical lobes expanding at 650 km/s, that are connected to each other near the central star ([6] and references therein). The equatorial plane orthogonal to the axis of the Homunculus contains ejected material expanding from the core at up to 1500 km/s. The bipolar shape of the Homunculus nebulae could be attributed to an equatorial disc composed of gas and dust, as in the case of butterfly nebulae [7], or due to the fact that Eta Carinae is a binary star system, as recently demonstrated [8]. Wikipedia is also reporting a theory that two small black holes may be at the center of each lobe, one of which is consuming the star.

The Homunculus from Eta Carinae is actually one of the most studied object, as the start itself. Eta Carinae is changing its brightness, and currently is classified as a luminous blue variable binary star. Wikipedia again reports the history of Eta Carinae brightness. In April 1843, the star reached its greatest apparent brightness and it was the second brightest star in the night-time sky after Sirius. About the time of its maximum brightness, it is highly probable that Eta Carinae created the Homunculus. The approximate distance of Eta Carinae is 7,500 light-years, that is quite away: it is nevertheless possible to distinguish in the nebula many structures with the size of about the diameter of the solar system [5]. The images show knots, lanes and radial streaks originated from the star.

**Homunculus imaged by the Hubble Space Telescope**
Many images of Homunculus can be seen, obtained from the Hubble Space Telescope (HST). In fact one of the early announcement about HST observations was on the resolution of individual clumps in the Homunculus, with a size of about ten times the size of the Solar System, obtained with the Wide Field and Planetary Camera. According to those observations, the nebula was considered as a thin and well defined shell of material, rather than a filled volume. Knots and filaments trace the locations of shock fronts within the nebula [9].

One of the best images is that proposed by Jon Morse, University of Colorado and NASA-HST, at [10], here reproduced as Fig.1.a. The image is among those in Ref.6: this paper has a remarkably description of what the Hubble Space Telescope revealed. The telescope displayed the complex network of bright cells and dusty lanes, lacing the surfaces of the bipolar lobes of the Homunculus. A debris field is expanding beyond the Homunculus. This region contains a set of high-velocity whiskers or streamers associated with the radially expanding lobes of the Homunculus and extend away from the central star like debris of an explosion. The HST observation discovered also an excess of UV light at the northwest of the central star in the vicinity of a burst of radio emission. This blue glow appears to emanate from the equatorial region between the bipolar lobes.

As reported in [6], using a combination of image processing techniques, the researchers created one among the highest resolution images of an extended object. The resulting picture is so detailed that, even the nebulas is about 7,500 light-years away, structures of about the diameter of our solar system can be distinguished. The Carina Nebula was observed in September 1995 with the Wide Field Planetary Camera 2 (WFPC2). Images were taken through red and ultraviolet filters [10].

Taking advantage of the spatial resolution of HST measurements, a two-dimensional map of the amount and position angle of the polarization across the Homunculus was proposed [11,12]. The data provide insight into the three-dimensional distribution of dust about the star and in the small-scale dust distribution on the lobes, which gives their cauliflower appearance.

The HST observation clearly confirms that the lobes are essentially hollow. One of the lobes is not a sphere, as it is possible to see from a "flask" edge on its surface [6]. An excess violet light escapes along the equatorial plane between the bipolar lobes. Apparently, there is relatively little dusty debris between the lobes and most of the blue light is able to escape. The lobes, on the other hand, contain large amounts of dust which preferentially absorb blue light, causing their reddish appearance.

Fig.1.a shows the image as it is from [10]. Applying a further image processing, the GIMP curve tool and GIMP brightness-contrast tool, we find 1.b and 1.c respectively. These two images are the best we can do with GIMP, without loosing too many details of the central region. We have a better view of the NW lobe and we see also some rays as whiskers originating from the central star. Fig.1.d has been prepared with another tool, AstroFracTool, to enhance the image edges, based on the use of the fractional gradient [13,14]. The image obtained with AstrFracTool was slightly adjusted with GIMP brightness-contrast tool. The final resolution of this image is better than 1.b and 1.c. Image 1.d shows that the NW lobe has the same cauliflower structure, with a protuberance resembling the flask shape of the SE lobe.

The reader can see that rays are originated from Eta Carinae and also from the two stars at top left corner of the image (see 1.d). Probably, among the observed long Homunculus' whiskers, there are some which are not properly represented or even artificially created, because of the point-spread function of the instrumentation.

AstroFracTool is useful to enhance the image edges, maintaining the image visibility. Pure edge detections can be easily obtained with other processing methods, such as the Sobel algorithm or the recently proposed dipole algorithm [15,16]: image 1.e is obtained from 1.a with the GIMP Sobel tool whereas image 1.f is built with the color dipoles algorithm [15]. These two images provide us a better view of what is told in Ref.6, that there is evidence for large-scale color gradients across the lobes of the nebula. It is clear that the core region near the central star contains more blue than the outer parts of the red lobes. Note the blue "fan" in the NW lobe. As Fig,1 shows, a further image processing seems to be useful in enhancing some specific features.

**Homunculus imaged by the Gemini South Telescope**
Fig.2.a shows Eta Carinae as imaged by the Gemini South telescope in Chile with the Near Infrared Coronagraphic Imager (NICI), which is using an adaptive optics to reduce the blurring effect coming from the turbulence in Earth's atmosphere. The original image is at [17]. As described in Ref. 18, this image shows the bipolar lobes of the Homunculus Nebula, visible with the never-before imaged "Little Homunculus Nebula" as a faint blue glow. The central star system appears as a dark spot due to an occulting disk used to eliminate the star's glare. The image is a composite one, obtained using three different infrared filters, the corresponding data are rendered with red, green and blue tones of pixels. The image was proposed at the 215th American Astronomical Society Meeting. It is, as the HST image, a high-resolution view of the complex structure of the glowing gas and dust surrounding the star.

The Gemini image is due to the research of John Martin and his team [19]. According to the researchers, the image displays a feature of the nebula never directly imaged before, the Little Homunculus, which is under the visible outer layer of the great outburst, corresponding to the Homunculus. In early 2007, Eta Carinae revealed new surprising features: the ground-based observations indicated that the star was rapidly decreasing in brightness. As reported in [20], a chaotic variation in brightness is possibly coming from viewing the star directly along the unstable boundary between low and high latitude winds. For the study of Eta Carinae, combined researches with Gemini South and HST were used to compare the spectra in late June 2007 [20].

As previously done on the HST image, we can try to apply a further image processing to the Gemini image. The use of GIMP curve and brightness-contrast tools were not able to resolve details: in fact, they strongly reduce the image quality. The use of AstroFracTool instead, is able to show many details. Images 2.b and 2.c were obtained with different fractional and visibility parameters ($\nu=0.8$, $\alpha=0.4$ and $\nu=1.0, \alpha=0.4$, respectively, see Refs.13 and 14 for the meaning of these parameters). Note the pattern formed by the debris of explosion between the lobes in image 2.c. The shape of the NW lobe looks different from the HST image but this is simply due to the fact that the top right corner of the original image is cut.

**Comparing images**
The shape of the lobes is the same as imaged from HST and from Gemini. The "fan" on the NW lobe is again visible in 2.b after enhancement (note the slight embossment effect of the algorithm). The detail of the laced structures of lobes is actually reduced in the Gemini image: nevertheless, the SE lobe has the "flask" edge on its surface, as clearly shown by image 2.b.

In the lower part of figure, in image 2.c, it is possible to note many rings, concentric with the star, probably due to spread function of the instrumentation. HTS and Gemini instruments have two different point-spread functions and then, after the processing deconvolution methods have been applied to the relative images, the two systems give us an imaging of the central whiskers with different features, long and straight in the HST imaging, as curly hair in the case of the Gemini imaging.

We could ask ourselves whether a quantitative comparison of structures shown in HST and Gemini images is possible or not. In fact, these images are probably obtained from data recorded during quite different periods of time, and the variations in brightness of the star, which are not negligible, as well as the motion of the nebula itself, deeply affect the final result of any comparison. Moreover, each instrument has its specific function affecting the final rendering.

It is not easy then to answer positively or negatively. Here, we show just a possibility based on the use of edge detection algorithms to find reference structures. First of all, we need to enhance knots and filamentary structures in the Gemini original image. The upper part of Fig.3 is obtained using the color dipole method to enhance the edge. In this case, the method is applied on the image corresponding to the green tones. In the lower part, we report an HST image, adapted from a figure in Ref.6. The reader can observe some knots that seem to correspond in both images encircled in red. Assuming these knots as reference points, we observe that several filamentary structures seems to correspond too. Note that in Fig.3 the dipole edges of the Gemini images are compared to the structures shown by the original image from HST, not with its dipole edges.

Fig.1 and Fig.2 separately show that a further image processing can be suitable to enhance specific details in the original images. For comparing the images, we have seen through Fig.3 that the edge detection algorithm is a good starting point to develop a successful method. In spite of the quite different appearance of images obtained with different instrumentation, the use of edge enhancement reveals some specific details that can be seen as reference points. After identifying some of these reference points, it seems more easy to recognize the structures passing from one image to the other..

**Note**
For the reader convenience, Figures 1, 2 and 3 of this paper are shown with a better resolution at http://staff.polito.it/amelia.sparavigna/Homunculus-images/

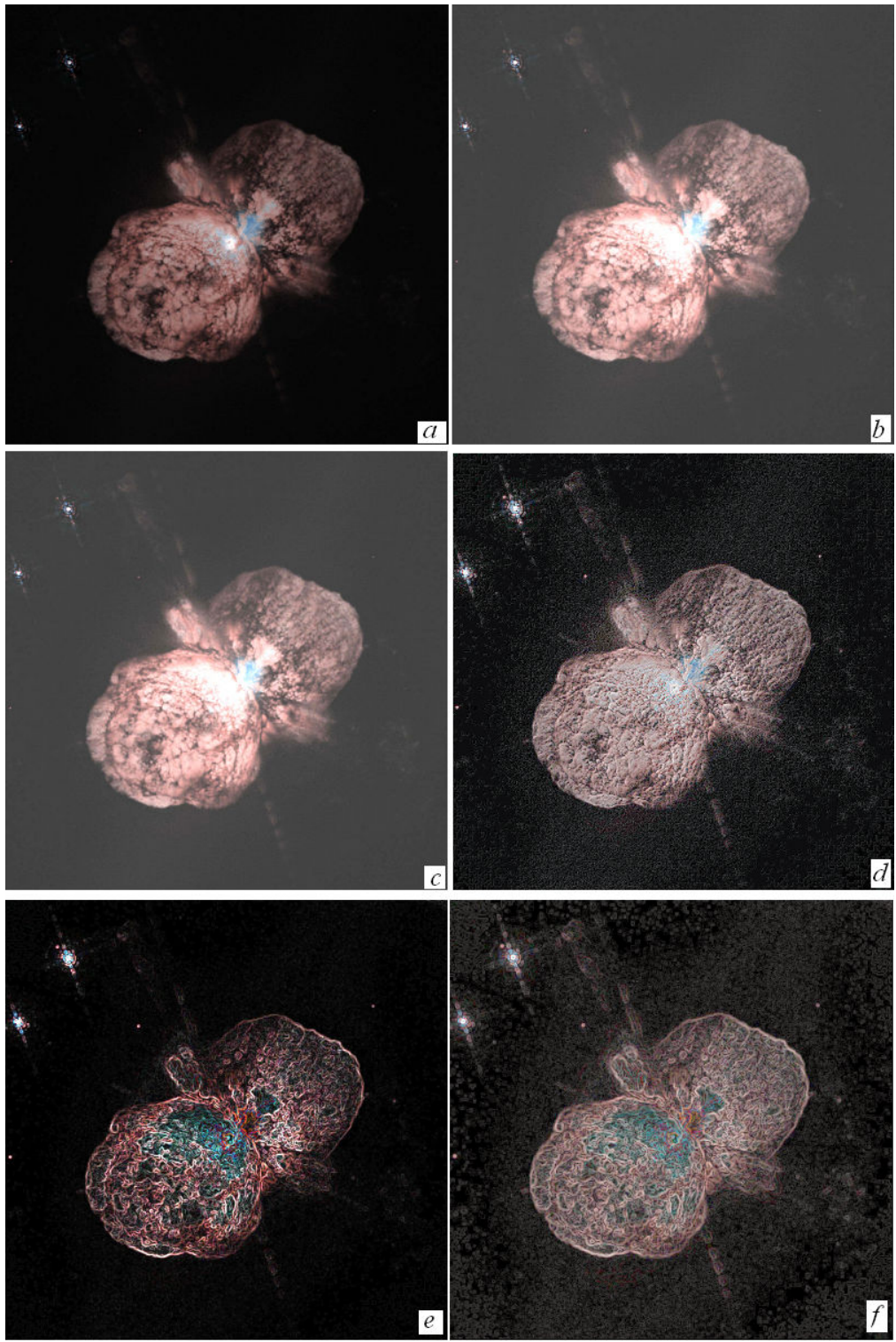

Fig.1 - 1.a shows the image as it is from ref.10. 1.b and 1.c are obtained after using the GIMP curve and brightness-contrast tools. These are the best results we can obtain, without loosing details of the central region. 1.d is obtained enhancing the image edges with AstroFracTool and GIMP brightness-contrast tool. The resolution of 1.d is better than that of 1.b and 1.c. Rays are originated from Eta Carinae and from the two stars at the top left corner. Among Homunculus' whiskers, there are some due to the point-spread function of the instrumentation. In the lower part of the figure, 1.e and 1.f are images obtained applying the GIMP Sobel tool and the colour dipole algorithm [15] respectively. Note that these two images reveal the large-scale colour gradients across the lobes of the nebula.

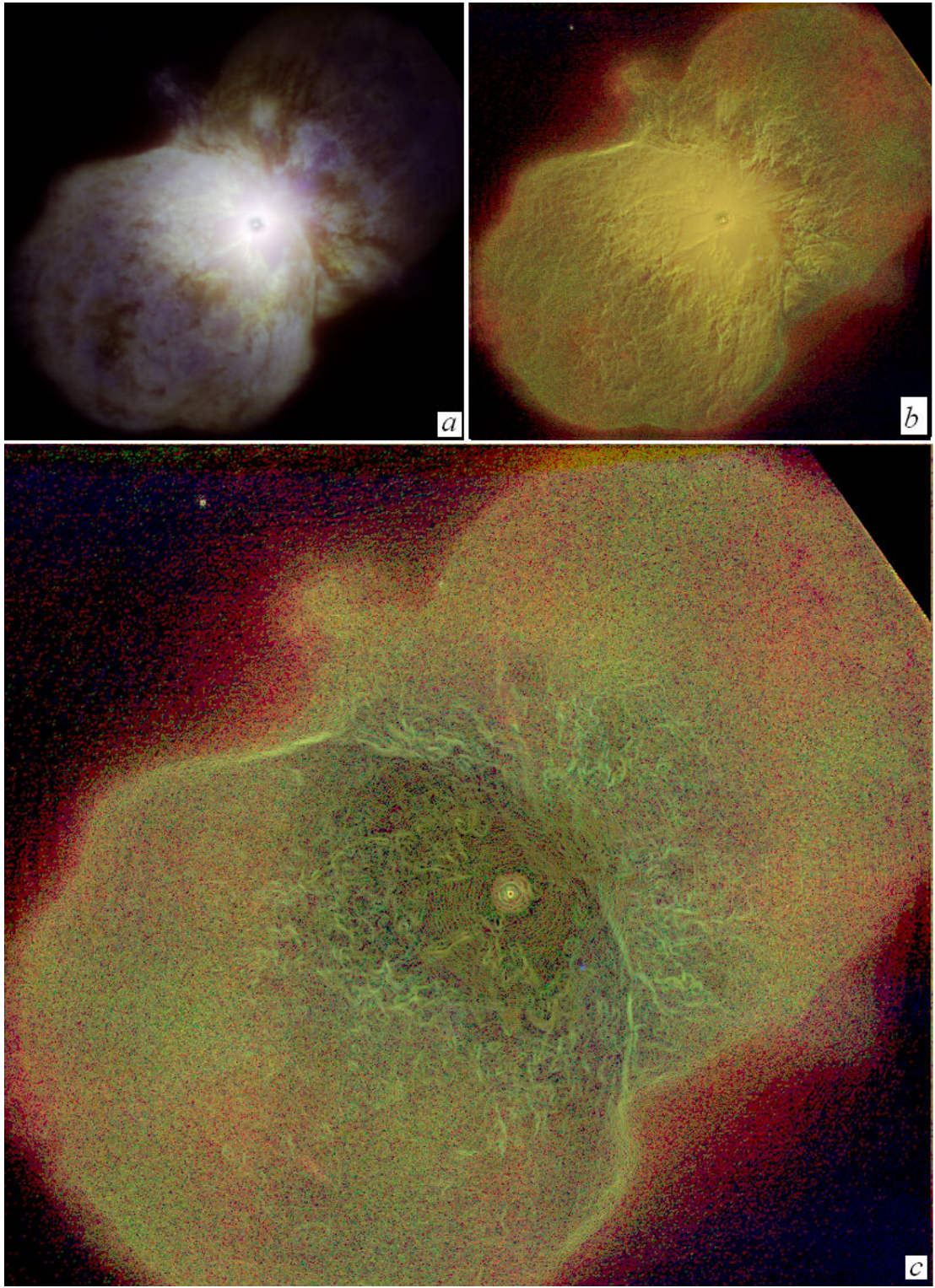

Fig.2 - 2.a shows the Homunculus imaged by the Gemini South Telescope in Chile with NICI. The image clearly displays the bipolar structure of the Homunculus Nebula, with inside the "Little Homunculus" as a faint blue glow. The image is a composite one obtained with three infrared filters, rendered in red, green and blue tones. 2.b and 2.c are obtained using AstroFracTool on the original image with two different parameter selections. Note that in 2.c we can see many rings, concentric with the star, probably due to the spread function of the instrumentation.

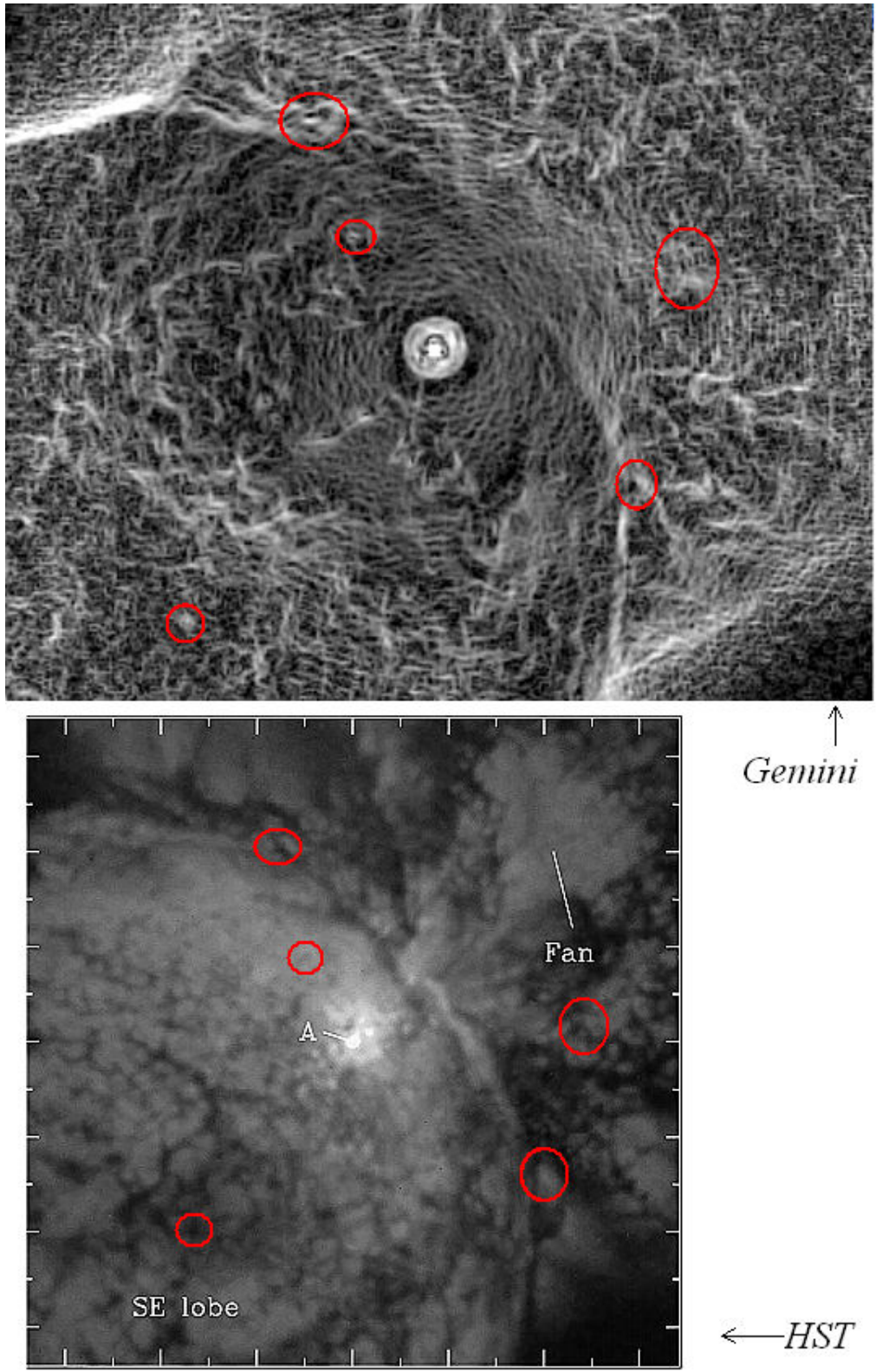

Fig.3 – Using the dipole algorithm on the green tones of the Gemini image, we can compare it with an image from HST (adapted from Fig.5 in Ref.6). In spite of the quite different appearance of the original images obtained with a different instrumentation, we can see many knots that seem to correspond in both images. These knots are encircled with red.